\documentclass[twocolumn,twocolappendix]{aastex631}

\newcommand{\beam}{$\theta_{\mbox{\scriptsize maj}}\times\theta_{\mbox{\scriptsize min}}$}

\shorttitle{JVLA Observations of ZZ Tau IRS}
\shortauthors{Hashimoto et al.}

\graphicspath{{./}{figures/}}

\begin{document}

\title{Grain Growth in the Dust Ring with Crescent around Very Low Mass Star ZZ~Tau~IRS with JVLA}

\correspondingauthor{Jun Hashimoto}
\email{jun.hashimto@nao.ac.jp}

\author[0000-0002-3053-3575]{Jun Hashimoto}
\affil{Astrobiology Center, National Institutes of Natural Sciences, 2-21-1 Osawa, Mitaka, Tokyo 181-8588, Japan}
\affil{Subaru Telescope, National Astronomical Observatory of Japan, Mitaka, Tokyo 181-8588, Japan}
\affil{Department of Astronomy, School of Science, Graduate University for Advanced Studies (SOKENDAI), Mitaka, Tokyo 181-8588, Japan}

\author[0000-0003-2300-2626]{Hauyu Baobab Liu}
\affiliation{Institute of Astronomy and Astrophysics, Academia Sinica, 11F of Astronomy-Mathematics Building, AS/NTU No.1, Sec. 4,
Roosevelt Rd, Taipei 10617, Taiwan, ROC}

\author[0000-0001-9290-7846]{Ruobing Dong}
\affil{Department of Physics \& Astronomy, University of Victoria, Victoria, BC, V8P 1A1, Canada}

\author[0000-0001-5830-3619]{Beibei Liu}
\affil{Institute for Astronomy, School of Physics, Zhejiang University, 38 Zheda Road, Hangzhou 310027, China}

\author{Takayuki Muto}
\affil{Division of Liberal Arts, Kogakuin University, 1-24-2, Nishi-Shinjuku, Shinjuku-ku, Tokyo 163-8677, Japan}

\begin{abstract}

The azimuthal asymmetries of dust rings in protoplanetary disks such as a crescent around young stars are often interpreted as dust traps, and thus as ideal locations for planetesimal and planet formations. Whether such dust traps effectively promote planetesimal formation in disks around very-low-mass stars (VLM; a mass of $\lesssim$0.2~$M_\sun$) is debatable, as the dynamical and grain growth timescales in such systems are long. To investigate grain growth in such systems, we studied the dust ring with crescent around the VLM star ZZ~Tau~IRS using the Karl G. Jansky Very Large Array (JVLA) at centimeter wavelengths. Significant signals were detected around ZZ~Tau~IRS. To estimate the maximum grain size ($a_{\rm max}$) in the crescent, we compared the observed spectral energy distribution (SED) with SEDs for various $a_{\rm max}$ values predicted by radiative transfer calculations. We found $a_{\rm max} \gtrsim$~1~mm and $\lesssim$~60~$\mu$m in the crescent and ring, respectively, though our modeling efforts rely on uncertain dust properties. Our results suggest that grain growth occurred in the ZZ~Tau~IRS disk, relative to sub-micron-sized interstellar medium. Planet formation in crescent with mm-sized pebbles might proceed more efficiently than in other regions with sub-millimeter-sized pebbles via pebble accretion scenarios.

\end{abstract}


\section{Introduction} \label{sec:intro}

The formation of planets around very low-mass (VLM) stars and brown dwarfs with masses ${\lesssim}0.2 \ M_\odot$ is one of the key questions in modern planet formation theory. The core accretion scenario predicts that only low-mass planets of up to a few Earth masses can form in such systems \citep{Payne2007a,Ormel2017a,liu2019a,Liu2020a,miguel2020a,burn2021}. 
When the planet becomes sufficiently massive, it opens a partial gas gap on the disk. 
According to the pebble-driven planet formation models  \citep{Ormel2010a,Lambrechts2012,liu2019a}, 
the generated gas pressure bump at the edge of the planetary gap acts to trap drifting pebbles inward. Therefore, the planet was isolated from pebble accretion \citep{lambrechts2014}. The corresponding mass is called the `pebble isolation mass', which scales approximately linearly with the mass of the central host \citep{liu2019a}. A typical isolation mass is a few Earth masses for VLM stars and Mars masses for brown dwarfs \citep{Liu2020a}.

Such low-mass planets have been observed around VLM stars and brown dwarfs (e.g., TRAPPIST-1 in \citealp{Gillon2016a}; Proxima~Centauri in \citealp{Anglada-Escude2016a}; and YZ~Cet in \citealp{Astudillo-Defru2017a}). Meanwhile, giant gas planets have also been reported (e.g., GJ 3512 in \citealp{morales2019a}; TWA~27 in \citealp{Chauvin2004}; 2MASS~J04414489$+$2301513 in \citealp{Todorov2010a}). The formation mechanism of these giant gas planets is unclear (i.e., cloud or disk fragmentation similar to the binary formation or core accretion in the protoplanetary disks for planets).

For massive planets, it is predicted that planet--disk interactions lead to gap opening \citep[e.g.,][]{lin1986}. Such ring/gap structures have been revealed in tens of protoplanetary disks around solar- and intermediate-mass stars \citep[e.g.,][]{andr2018,long2018a} and VLM stars \citep[e.g.,][]{Pinilla2018a,Pinilla2021a,Kurtovic2021a,Hashimoto2021zztauirs,Hashimoto2021sz84} by Atacama Large Millimeter/submillimeter Array (ALMA). Hydrodynamic simulations show that a giant planet with a minimum mass of 1.4~$M_{\rm Jup}$ can explain the ring/gap structure around the VLM star CIDA~1 \citep{Curone2022arXiv}. 

In addition to gaps, crescent structures have been observed in disks \citep[e.g.,][]{vandermarel20a}. These are thought to be dust traps caused by long-lived anticyclonic vortices \citep[e.g.,][]{Barge1995,birn13}, which may form at gap edges opened by planets \citep[e.g.,][]{raettig2015}. A high concentration of dust in crescent structures may enhance grain growth, making them ideal locations for planetesimal formation.

Because key planet formation processes, such as streaming instability and pebble accretion, depend crucially on the size of dust particles \citep{Liu2020a,Drazkowska2022}, inferring the maximum grain size in protoplanetary disks is extremely important. There are two well-established methods to constrain dust sizes based on millimeter/centimeter continuum emission. One is to observe polarized millimeter/centimeter continuum emissions due to scattering \citep[e.g.,][]{Kataoka2015a}. The polarization fraction at the observation wavelength $\lambda$ reaches a maximum when the dust grains grow to a maximum size of $a_{\rm max} \sim \lambda/2\pi$. To probe centimeter-sized pebbles, longer-wavelength observations at $\lambda \gtrsim$10~cm are suitable to probe centimeter-sized pebbles. Another avenue is to construct the spectral energy distribution (SED; e.g., \citealp{Liu2019b}) because grains efficiently emit thermal radiation at a wavelength similar to their sizes \citep[e.g.,][]{Draine2006}. 

Centimeter emissions have been detected in rings and crescents at tens of au around T-Tauri and intermediate-mass stars (e.g., HD~142527 in \citealt{casassus+2015}; Oph~IRS~48 in \citealt{vandermarel+2015}; MWC~758 in \citealt{casassus2019a}; LkCa~15 in \citealp{isella2014}). However, sub-structures around VLM stars at centimeter wavelengths have not yet been well explored. In this study, we report the JVLA observations of ZZ~Tau~IRS at Ka ($\nu=$~33~GHz: $\lambda=$~9.1~mm), Ku ($\nu=$~15~GHz: $\lambda=$~20~cm), and X ($\nu=$~10~GHz: $\lambda=$~30~cm) bands as a pilot study of grain growth in a ring with crescent around VLM stars. 

The VLM star ZZ~Tau~IRS with a mass of $\sim$0.1-0.2~$M_\sun$ \citep{Andrews2013} is located at a distance of 130.7~pc in the Taurus star forming region \citep{akeson2019}. A dust disk with a mass of $\sim$0.3~$M_{\rm Jup}$ \citep{Andrews2013} is the most massive among VLM stars \citep[e.g.,][]{ansdell2017a}. The ring structure at $r\sim$~60~au is also the largest, and shows a crescent with a azimuthal contrast of 1.4 at $\lambda =$~0.9~mm \citep{Hashimoto2021zztauirs}. To date, only ZZ~Tau~IRS is known to possess the ring with crescent among VLM stars. The inner disk may be misaligned to the outer ring with a misaligned angle of 30$\degr$ \citep{white2004,furlan2011,Hashimoto2021zztauirs}, suggesting the presence of an inclined Jovian mass planet in the gap \citep[e.g.,][]{zhu2019a}. 

\section{Observations and data reduction} \label{sec:obs}

The JVLA observations were conducted as part of the program ID 21B-043 (PI: J. Hashimoto), as summarized in Table~\ref{tab:obs}. The data were calibrated using the Common Astronomy Software Applications (CASA) package \citep{mcmu07} following the calibration scripts provided by JVLA. We carefully inspected the calibrated data and performed additional flagging of the data with high and low amplitudes. Subsequently, we reran the calibration scripts. We did not perform self-calibration of the visibilities owing to the weak emission from the object. The stellar position was shifted by correcting for proper motion (10.675, $-$14.991)~mas/yr \citep{gaia2020a}.

\begin{deluxetable}{lc}[htbp]
\tablewidth{0pt} 
%\tablenum{1}
\tablecaption{JVLA observations\label{tab:obs}}
\tablehead{
\colhead{Observations}      & \colhead{}    
}
\startdata
Observing date (UT)         & 2021.Oct.2 and 11                \\
Project code                & 21B-043 (PI: J. Hashimoto)       \\
Central frequency (GHz)     & 33 (Ka), 15 (Ku), 10 (X) \\
Continuum band width (GHz)  & 8 (Ka), 6 (Ku), 4 (X) \\
Time on source (min)        & 81.9 (Ka), 10 (Ku), 10 (X)                 \\
Number of antennas          & 27 (Oct.2), 26 (Oct.10)                       \\
Baseline lengths (km)       & 0.243 to 11.1                  \\
Bandpass calibrator         & J0319$+$4130                     \\
Flux calibrator             & 3C147                     \\
Phase calibrator            & J0403$+$2600                   \\
\enddata
\end{deluxetable}

We performed zeroth order (i.e., nterm$=$1) multi-frequency synthesis imaging \citep{Cornwell2008MultiscaleClean,Rau2011MultiscaleClean} using the CASA {\tt tclean} task, as summarized in Table~\ref{tab:image} (top). For a fair comparison between these JVLA observations and the ALMA 338.8 GHz continuum image published in \citet{Hashimoto2021zztauirs}, we also produced continuum images with similar beam sizes, as summarized in Table~\ref{tab:image} (bottom). The images with similar beam sizes and the aforementioned ALMA 338.8 GHz images were smoothed to a 0$\farcs$53 circular beam before we measured the integrated flux density (Section \ref{sec:result}).
For clarity, we refer to these smoothed JVLA images as the 10, 15, and 33 GHz images in the following discussion.

Finally, we also produced Briggs Robust$=$0 weighed images for the three JVLA bands to constrain the radio emission inside of the dust cavity of ZZ~Tau~IRS, which is likely contributed by free-free emission, synchrotron emission, or their combination (Section \ref{sec:amax}).

\begin{deluxetable*}{lccc}[htbp]
%\tabletypesize{\footnotesize}
\tablewidth{0pt} 
%\tablenum{2}
\tablecaption{Imaging parameters \label{tab:image}}
\tablehead{
\colhead{}      & \colhead{X band (10 GHz)} & \colhead{Ku band (15 GHz)} & \colhead{Ka band (33 GHz)}    
}
%\colnumbers
\startdata
\multicolumn{4}{c}{For top and middle panels in Figure~\ref{fig:JVLAimg} with no smoothing.} \\
Robust clean parameter      & 0.0  & 2.0  & 2.0  \\
Beam shape                  & 0$\farcs$52$\times$0$\farcs$51; P.A.$=24^{\circ}$& 0$\farcs$51$\times$0$\farcs$47; P.A.$=47^{\circ}$ & 0$\farcs$24$\times$0$\farcs$20; P.A.$=47^{\circ}$ \\
r.m.s. noise ($\mu$Jy/beam) & 7.3 & 4.1 & 5.9 \\ \hline
\multicolumn{4}{c}{For bottom panels in Figure~\ref{fig:JVLAimg} with smoothing to a 0\farcs53 circular beam.} \\
Robust clean parameter      & 0.0  & 2.0  & 2.0  \\
Beam shape                  & 0$\farcs$53$\times$0$\farcs$53 & 0$\farcs$53$\times$0$\farcs$53 & 0$\farcs$53$\times$0$\farcs$53 \\
r.m.s. noise ($\mu$Jy/beam) & 7.3 & 4.3 & 8.7 
\enddata
\end{deluxetable*}

\subsection{Images and Flux Assessments} \label{sec:result}

\begin{figure*}[htbp]
         \includegraphics[width=6cm]{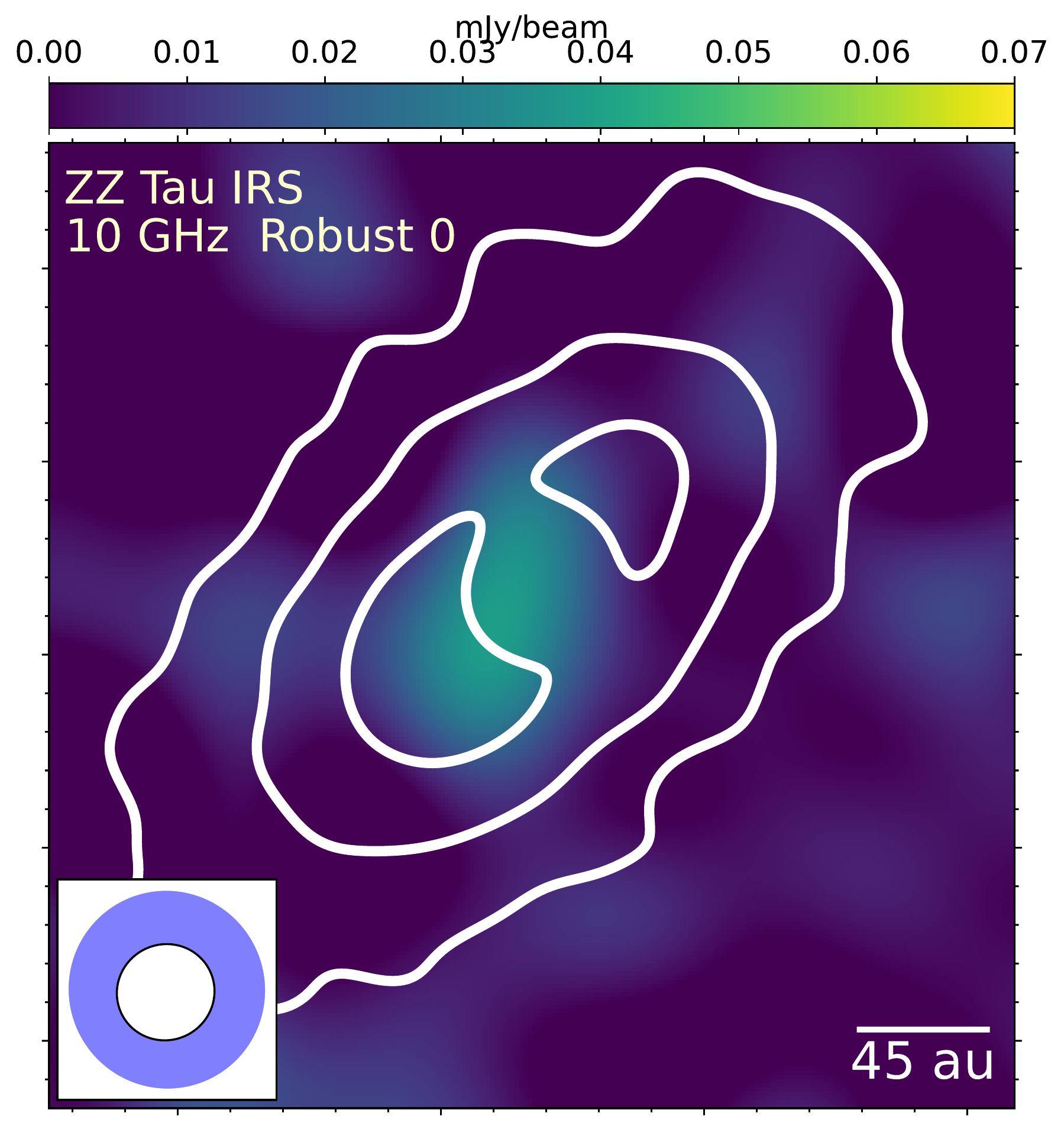} %&
         \includegraphics[width=6cm]{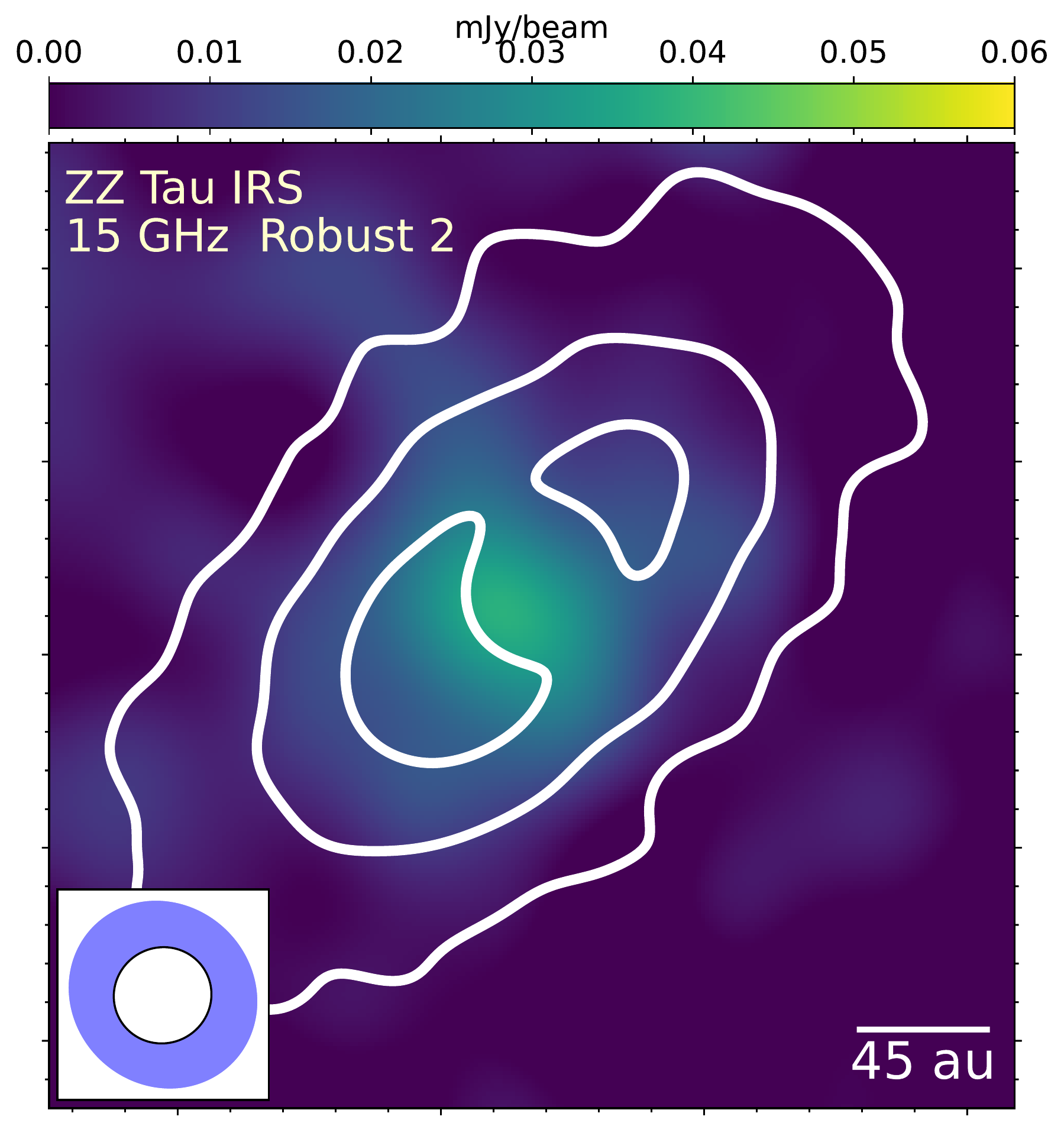} %&  
         \includegraphics[width=6cm]{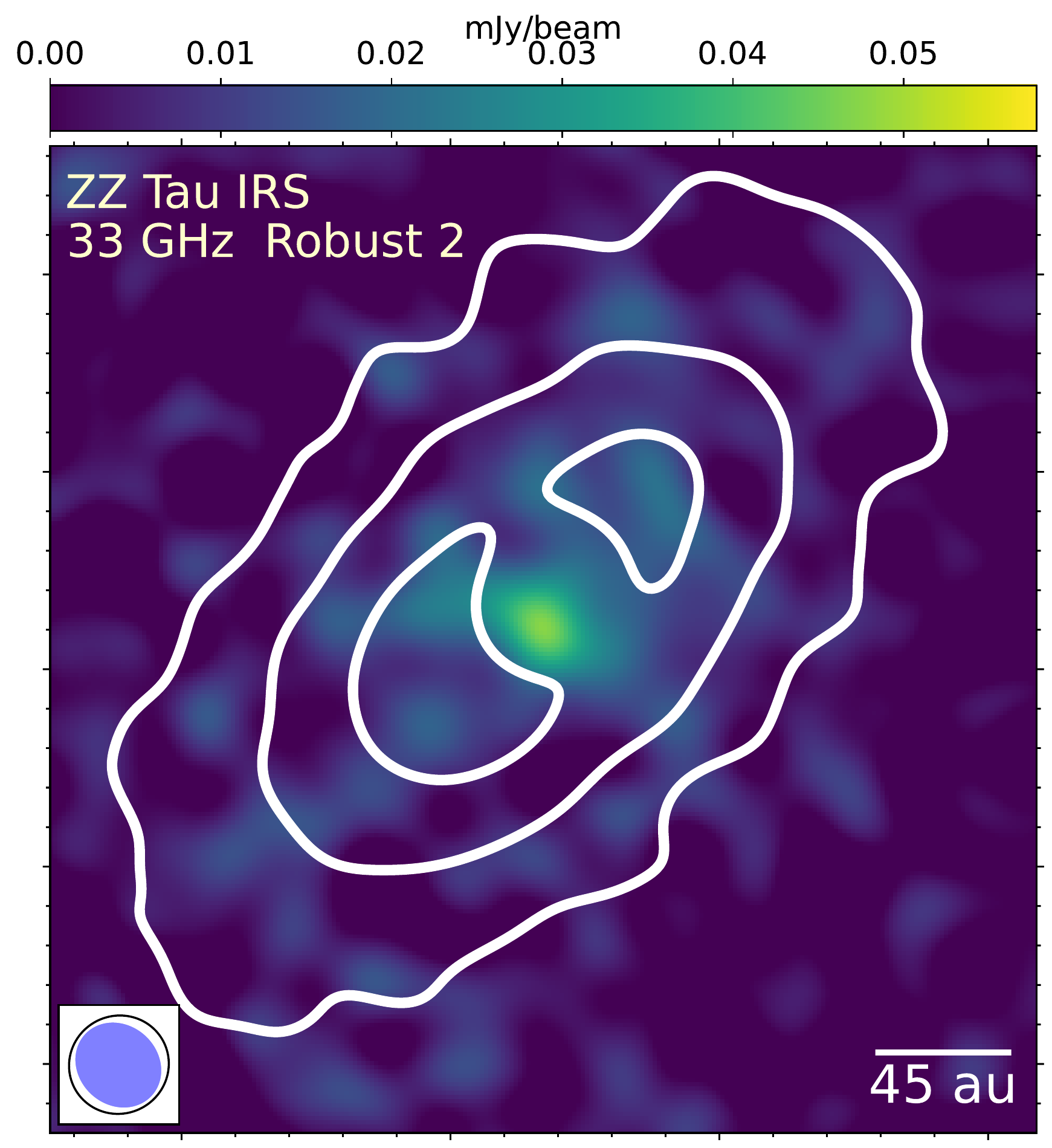} %\\
         \includegraphics[width=6cm]{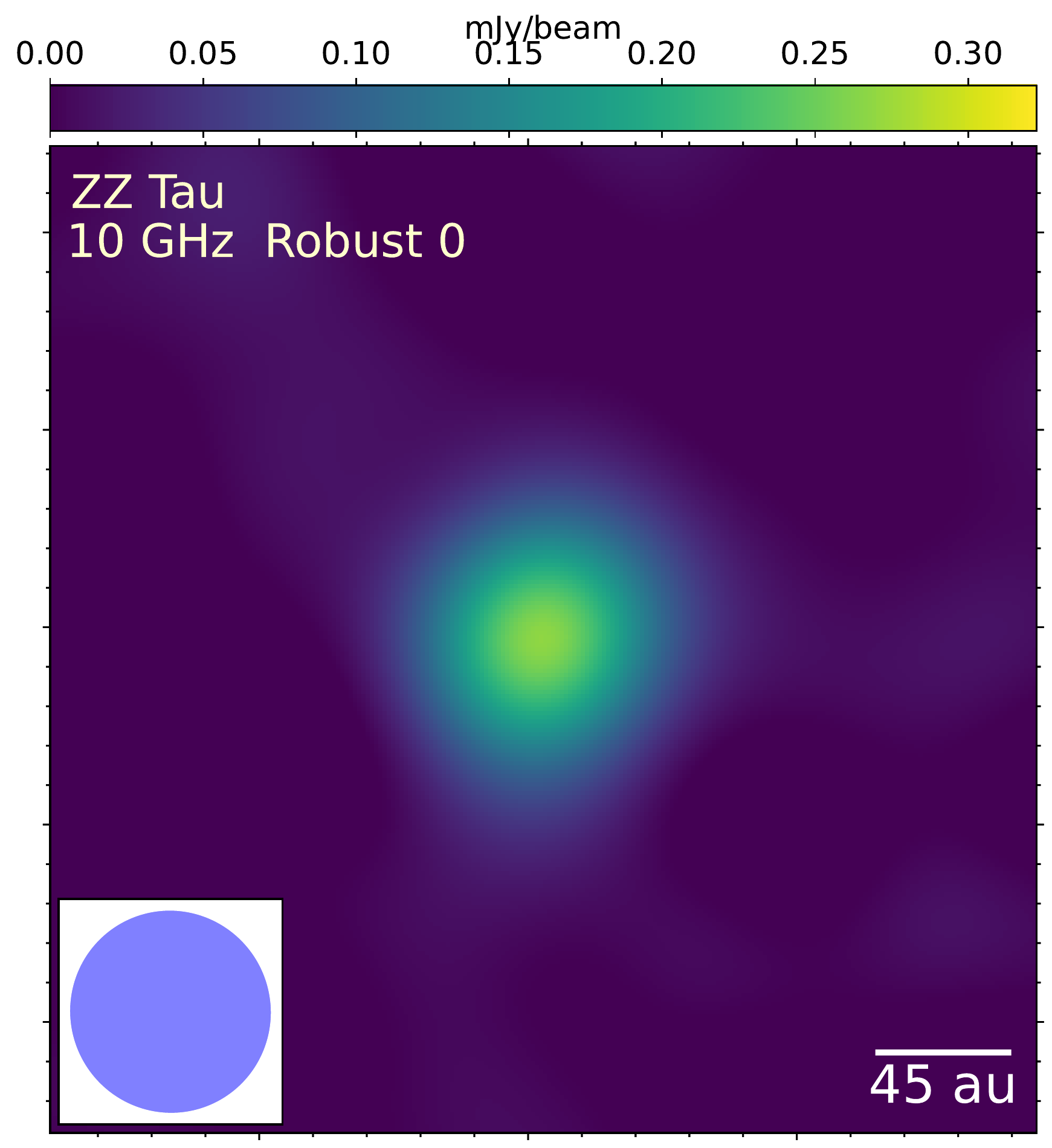} %&  
         \includegraphics[width=6cm]{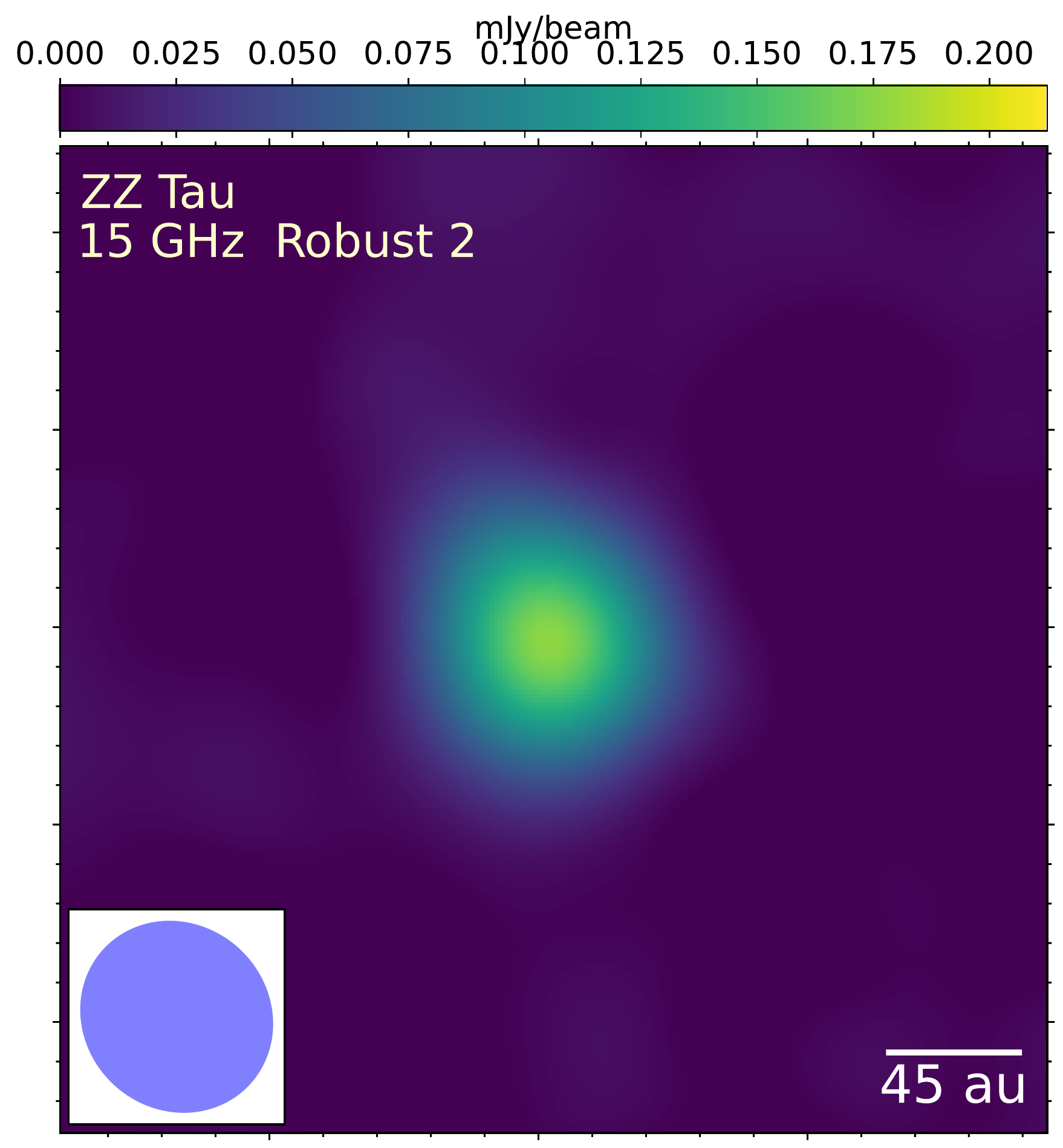} %&  
         \includegraphics[width=6cm]{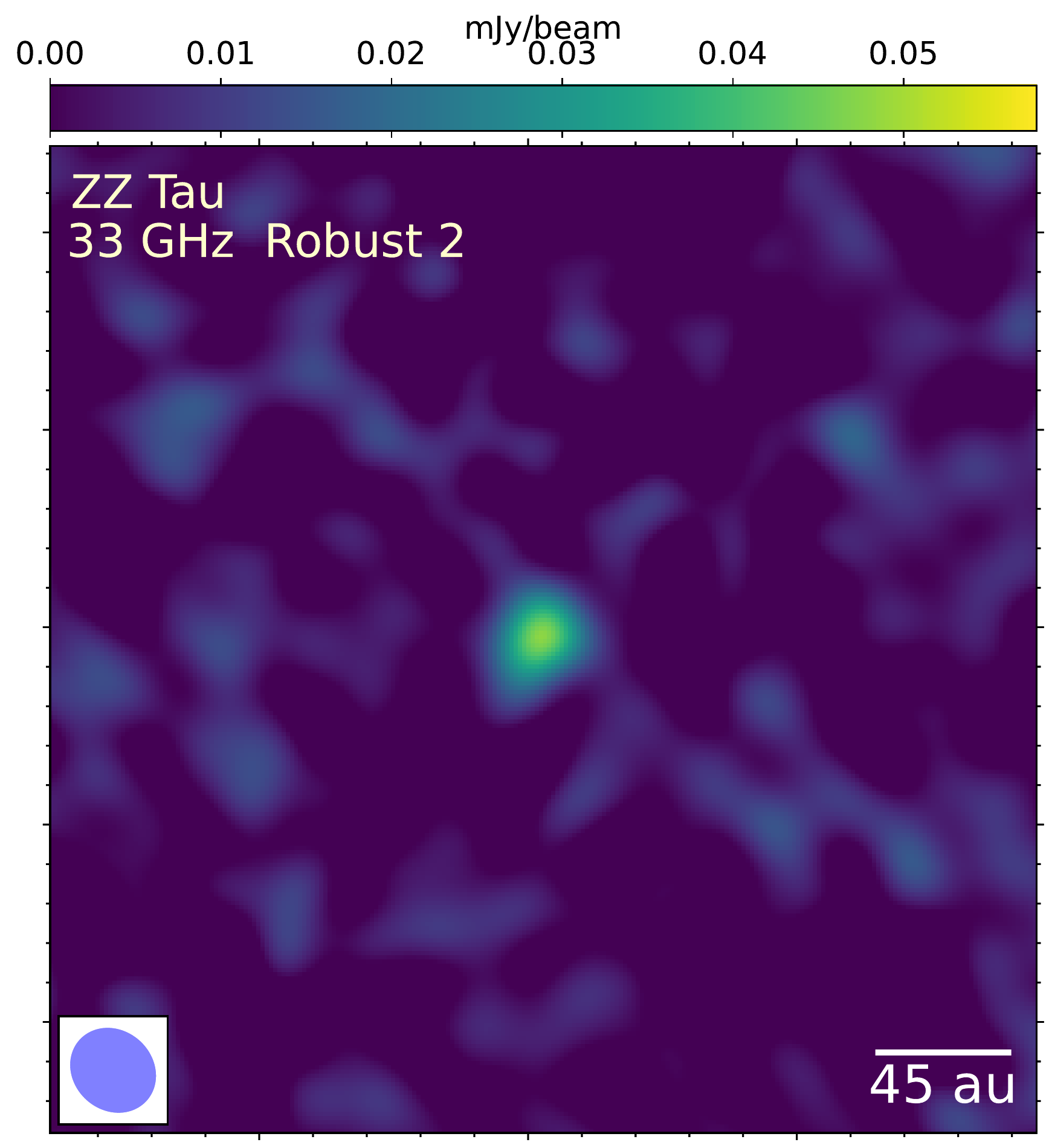} %\\
         \includegraphics[width=6cm]{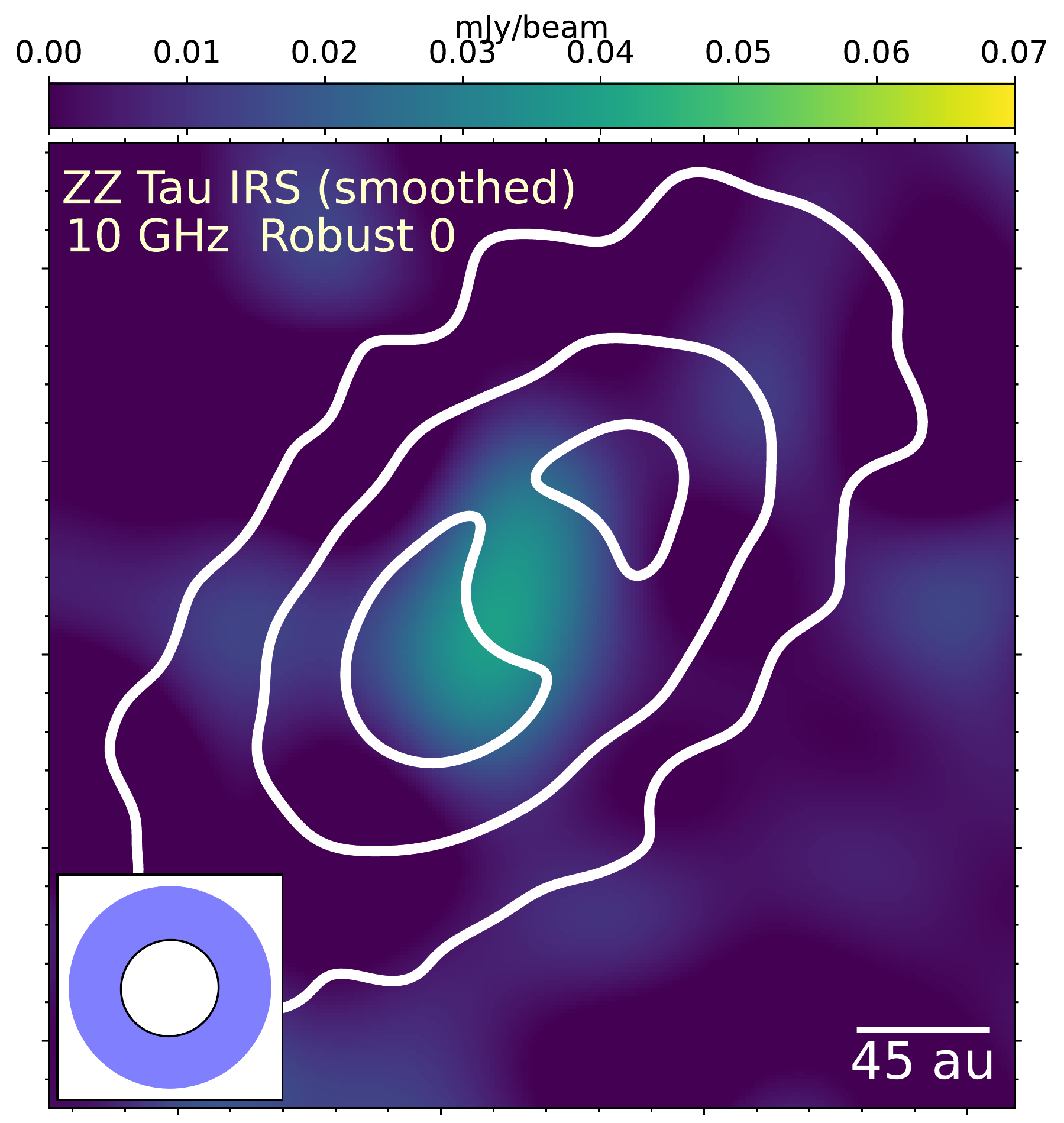} %&  
         \includegraphics[width=6cm]{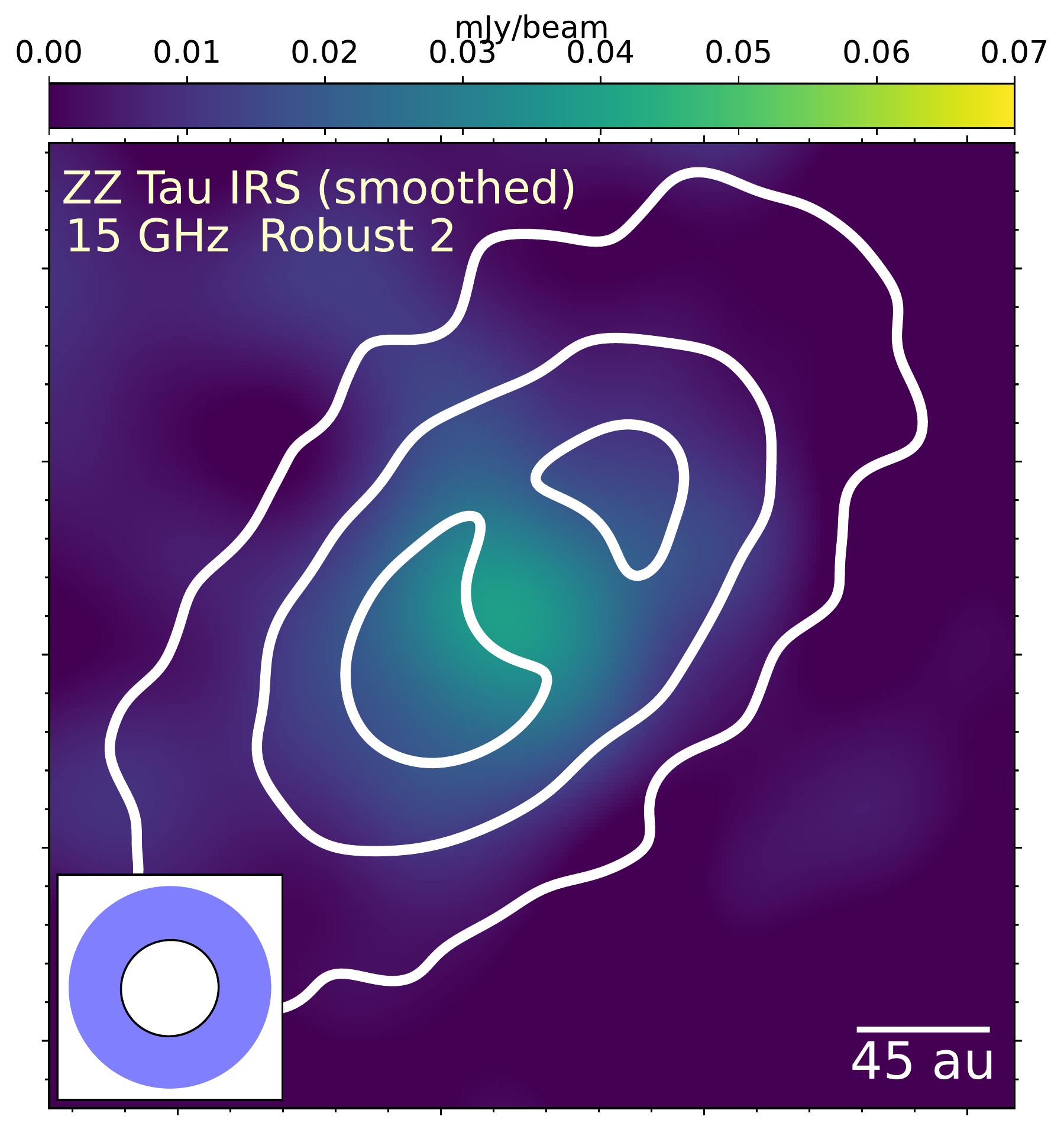} %&  
         \includegraphics[width=6cm]{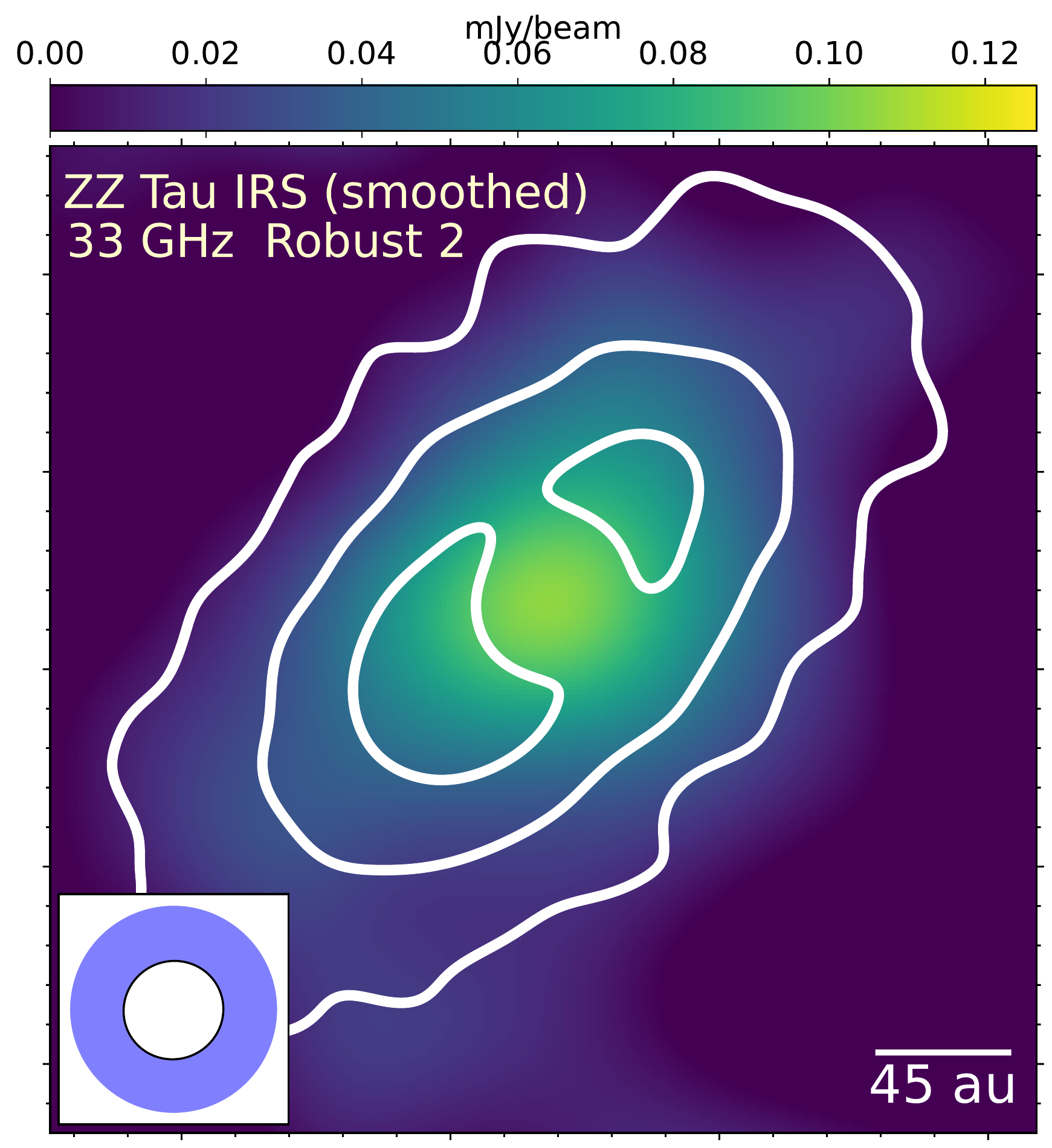} %\\
    \caption{
    JVLA images (color) overplotted with the ALMA 338.8 GHz image (contour; \beam=0$\farcs$25$\times$0$\farcs$24; P.A.=$-62^{\circ}$; \citealt{Hashimoto2021zztauirs}). Contour levels are 0.81 mJy\,beam$^{-1}$ (3~$\sigma$) $\times$[1, 5, 15].  
    From left to right the columns show the X band (8--12 GHz), Ku band (12--18 GHz), and Ka band (29--37 GHz) images, respectively. The top row shows the Robust$=0$ weighted X band image and the Robust$=2$ weighted Ku and Ka band images of ZZ~Tau~IRS. The middle row shows the images of ZZ~Tau that were produced together with those in the top row. The bottom row shows the smoothed, 0$\farcs$53 resolution images (refer to Section \ref{sec:obs}). The synthesized beams of the JVLA observations are shown by the blue ellipses in the lower left corner; the synthesized beam of the ALMA 338.8 image is shown by the white ellipses. 
    }
    \label{fig:JVLAimg}
\end{figure*}

Figure \ref{fig:JVLAimg} shows 10, 15, and 33~GHz JVLA images.
They are compared with the 338.8 GHz image previously published by \citet{Hashimoto2021zztauirs}.
The 338.8 GHz image resolved a spatially extended, dust ring with crescent around its inner cavity.
Around 35$''$ northwest of ZZ~Tau~IRS, our JVLA observations also detected another source, ZZ~Tau, which is not our present focus and will only be briefly discussed.

The 10 GHz image of ZZ~Tau~IRS resolved a spatially compact object.
The peak of the 10 GHz emission is inside the dust cavity.
There might be a weak contribution from the dust ring of ZZ~Tau~IRS (for further discussion, refer to Section \ref{sec:amax}), which we presently cannot robustly confirm owing to the poor angular resolution and high noise level of the 10 GHz image.

\begin{figure*}[htbp]
\begin{centering}
\includegraphics[clip,width=\linewidth]{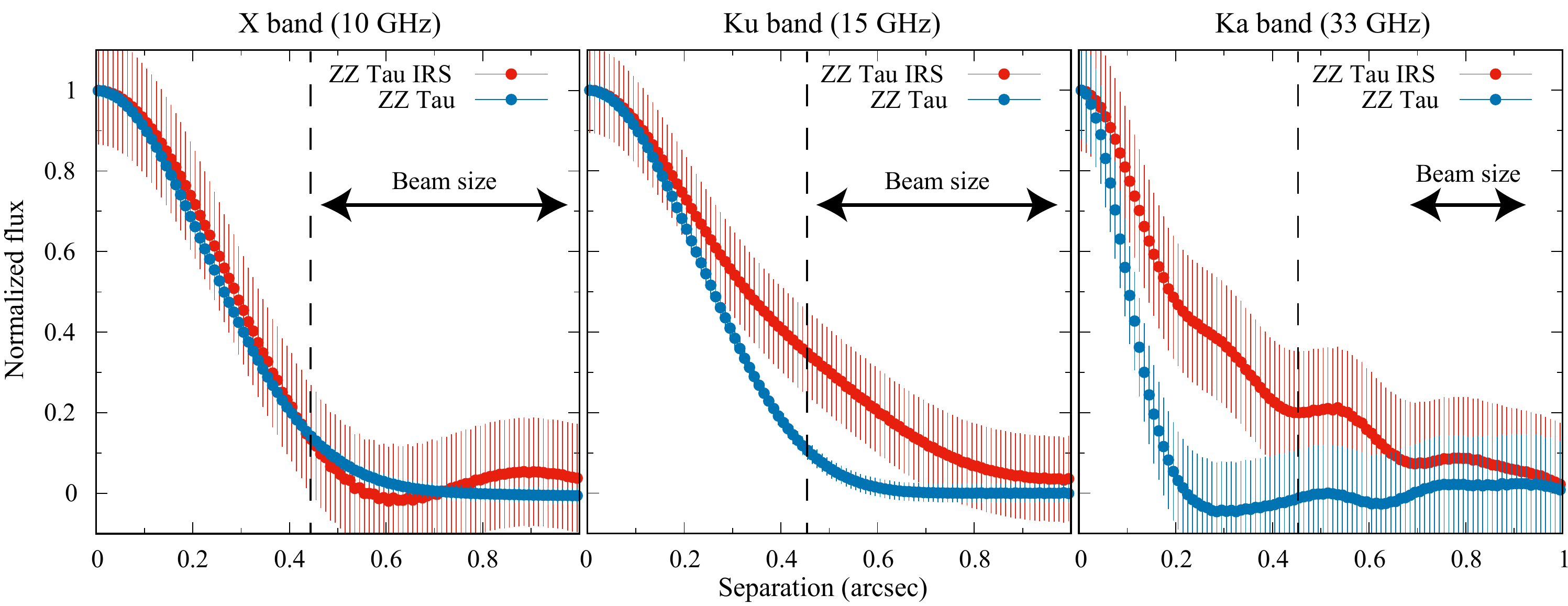}
\end{centering}
\caption{Radial profiles of ZZ~Tau~IRS and ZZ~Tau in top and middle rows in Figure~\ref{fig:JVLAimg}, respectively. Vertical dashed line represents the ring location detected in ALMA band-7 dust continuum image \citep{Hashimoto2021zztauirs}. The emission of ZZ~Tau~IRS at 15 and 33~GHz is more spatially extended than those of ZZ~Tau, whereas both ZZ~Tau~IRS and ZZ~Tau are not spatially resolved at 10~GHz.
} 
\label{fig:radprof}
\end{figure*}

The 15 GHz and 33 GHz images of ZZ~Tau~IRS both resolve a compact source inside the dust cavity and a low-intensity halo around the radii of the dust ring, as shown in the radial profiles in Figure~\ref{fig:radprof}. This low-intensity 15 GHz and 33 GHz emission halo cannot be attributed to calibration errors or imaging artifacts for the following reasons.
First, the allowable weather conditions for the 33 GHz observations correspond to extraordinarily good weather conditions for the 15 GHz observations.
Because we considered the observations at all three bands in the same scheduling blocks, it is unlikely that the lower-frequency observations were subject to similar calibration errors as the 33 GHz observations presented similar artifacts.
Second, the 15 GHz and 33 GHz images of ZZ~Tau resolved only a compact source.
If the low-intensity halo around ZZ~Tau~IRS is due to imaging or calibration artifacts, it is difficult to understand why ZZ~Tau is not subject to similar artifacts.

When we smoothed the 10, 15, and 33 GHz images of ZZ~Tau~IRS to a 0$\farcs$53 resolution (refer to the bottom row of Figure \ref{fig:JVLAimg}), the 10 GHz emission remains as a compact source inside the dust cavity; the 15 GHz emission is detected at $\gtrsim$3-4~$\sigma$ out of the radii of the 45 AU dust ring\footnote{While \citet{Hashimoto2021zztauirs} reported that the ring location is $r=$~58~au based on visibility analyses, the radial profile of the dust continuum image shows that the peak location is $r\sim$40-50~au in their Figure~1.}; the 33 GHz emission is detected at $\gtrsim$3-4~$\sigma$ over the region approximately as spatially extended as the 338.8 GHz emission.

In summary, we consider that the 15 GHz and 33 GHz images detected the radio counterparts of the dust ring of ZZ~Tau~IRS.
These images and the 10 GHz image additionally detected some emissions inside the dust cavity, which may be attributed to the ionized gas that is closely associated with the host protostar.
A deeper discussion of their emission mechanisms is presented in the following section.

\section{Constraints on Maximum Grain Sizes} \label{sec:amax}

In the following analyses, we constrained the dust sizes in the ring and crescent in the ZZ~Tau~IRS disk. The basic physics employed to achieve this goal is the dependence of dust opacity on dust size. Given the same temperature and surface density, dust of different sizes produces different amounts of thermal emissions at various wavelengths \citep[e.g.,][]{Draine2006}. By studying the SED at each location on the disk and comparing it with dust emission models, it is possible to constrain the dust size along with other properties, such as surface density and temperature \citep[e.g.,][]{Liu2019b}.  

The measured flux at each location has contributions from both the dust and ionized gas. The latter dominates at lower frequencies (i.e., $\lesssim$~10~GHz, e.g., \citealp{zapa17}). Dust grains emit most strongly at wavelengths similar to their size.
Thus, cm-sized large dust strongly emitting at JVLA wavelengths can have its emission contaminated by ionized gas emission, while the dust of smaller sizes experiences these effects less. To constrain the size of large dust grains by fitting the SED, the dust thermal emission in the total flux must be separated from the ionized gas emission. The latter is expected to be close to the star and localizable in high-resolution observations. However, it is challenging to separate thermal dust emissions from free-free emissions in our observations because of the limited angular resolution. 

Therefore, we aim to obtain the lower limit of the maximum dust grain size $a_{\mbox{\scriptsize max}}$ for a larger $a_{\mbox{\scriptsize max}}$ with $\gtrsim$~1~mm. However, dust grains with small sizes (e.g., $a_{\mbox{\scriptsize max}} \lesssim$~100~$\mu$m) can be directly estimated with sub-millimeter emissions. A lower limit on $a_{\mbox{\scriptsize max}}$ for larger dust grains at each location can be achieved by discounting the local long-wavelength emissions from ionized gas as much as possible (Section~\ref{sub:fluxdensity}), and the remaining long-wavelength emission comes from dust, enabling dust size constraints (Section~\ref{sub:model}). 

\subsection{Contributions to measured flux densities}\label{sub:fluxdensity}

\begin{figure*}[htbp]
    \includegraphics[width=19cm]{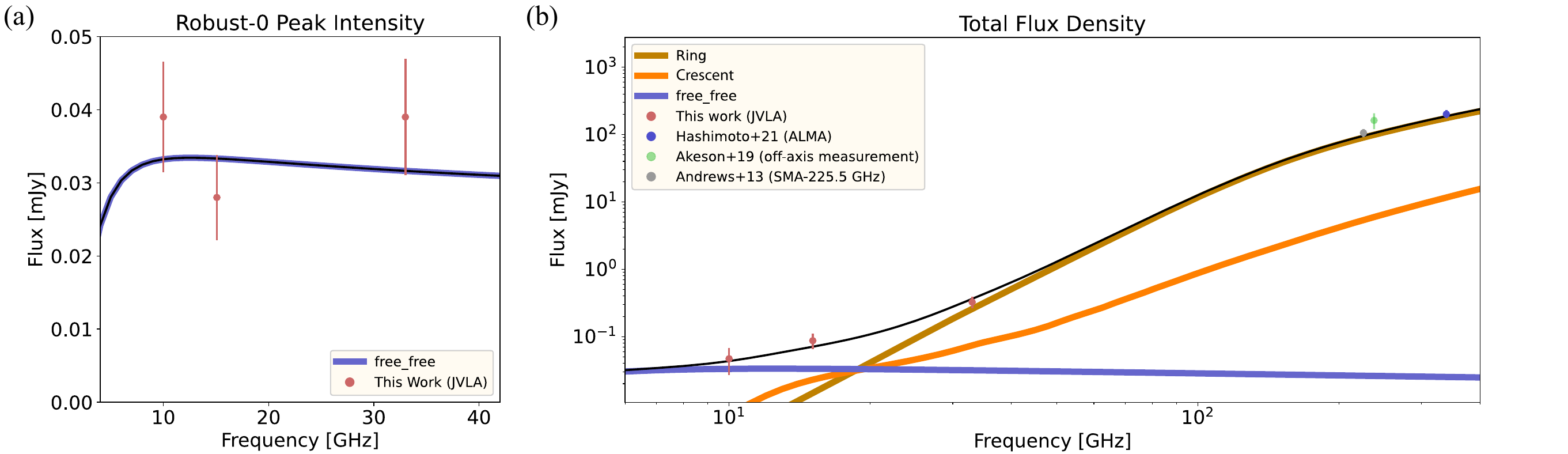}
    \caption{
    Observed flux densities (dots) from ZZ~Tau~IRS and our models to interpret them. Panel (a) shows the flux densities measured at the stellar position (i.e., the {\it star} column in Table \ref{tab:fluxdensity}). Panel (b) shows the integrated flux densities (i.e., the {\it Integrated} column in Table \ref{tab:fluxdensity}). In panel (b), the brown and orange lines represent the two dust emission components in our model (i.e., {\it Ring} and {\it Crescent}; refer to Table \ref{tab:dustmodel}), respectively; the dark-blue line represents the free-free emission component. The black line represents the total flux density. In panel (a), we only compare the measurements with the free-free emission given that the measurements were drawn from a region that is inside the dust cavity.
    We note that the off-axis ALMA measurement (green dot) was made well outside of the ALMA primary beam, therefore, it is not reliable. 
    }
    \vspace{0.5cm}
    \label{fig:sedmodel}
\end{figure*}

We assume that the emission of ionized gas (e.g., free-free emission or gyrosynchrotron emission) is limited to a small area around the host protostar and can only dominate the flux density at low frequencies, that is, 10, 15, and 33~GHz in our observations. While it is difficult to separate the emission into thermal and non-thermal components, we can constrain the maximum possible flux density (i.e., upper limit) of the ionized gas emission by reading the peak intensities at the stellar location in the robust $=0$ weighted images at these three bands, as summarized in Table \ref{tab:fluxdensity} and panel (a) of Figure \ref{fig:sedmodel}. The use of a smaller robust parameter yielded an extremely low signal-to-noise ratio for our measurement. We interpret these measurements as free-free emissions. After subtracting this free-free component, the residual flux density was modeled as the dust thermal emission in Section~\ref{sub:model}. 

We constrain the flux densities of the dust and ionized gas emission from the entire disk from the spatially smoothed, 0$\farcs$53 resolution images at all frequencies (Section \ref{sec:obs}; refer also to the bottom row of Figure \ref{fig:JVLAimg}) by integrating over the region that is above the 2-$\sigma$ contour of the 0$\farcs$53 resolution 338.8 GHz image. In addition, we quoted spatially unresolved SMA measurements at 225.5~GHz from \citet{Andrews2013}.

Finally, we found that ZZ~Tau~IRS was detected in previous ALMA 237.5 GHz observations (\citealt{akeson2019}; \beam$=$1$\farcs$02$\times$0$\farcs$54; P.A.$=47^{\circ}$), although it was well outside the primary beam.
To measure the flux density, we first smoothed the aforementioned 338.8 GHz image further to a synthesized beam \beam$=$1$\farcs$02$\times$0$\farcs$54; P.A.$=47^{\circ}$, and then integrated the flux densities of the 237.5 and 338.8 GHz images over the region that is above the 2-$\sigma$ contour of this further smoothed 338.8 GHz image.
This measurement of the 338.8 GHz flux density is 2\% lower than that measured from the 0$\farcs$53 resolution image, which can be safely attributed to thermal noise.
From this result, we also observe that missing short spacing is likely negligible at 338.8 GHz; missing short spacing is less likely to be important in our JVLA observations given that the maximum recoverable angular scales are larger, and dust emission is spatially more compact at longer wavelengths.
The flux density at 237.5 GHz was corrected for a nominal primary beam attenuation factor of 0.00855 (approximated with a Gaussian response function). 
We note that some pointing errors of the antennae can lead to large uncertainties in the primary beam attenuation factor.
The integrated flux densities are summarized in Table \ref{tab:fluxdensity} and panel (b) of Figure \ref{fig:sedmodel}.

\begin{deluxetable}{lcc}[htbp]
\tablecaption{
Flux densities in ZZ~Tau~IRS\label{tab:fluxdensity}
}
%\tablewidth{8cm}
\tabletypesize{\scriptsize}
\tablehead{
\colhead{Frequency} &
\colhead{Star\tablenotemark{a}} &
\colhead{Integrated\tablenotemark{b}} \\
\colhead{(GHz)} &
\colhead{(mJy)} &
\colhead{(mJy)} 
} 
\startdata
10 & 0.039$\pm$0.007 & 0.047$\pm$0.019 \\
15 & 0.028$\pm$0.006 & 0.087$\pm$0.019\\
33 & 0.039$\pm$0.008 & 0.325$\pm$0.052\\
225.5\tablenotemark{c} & $\cdots$ & 106$\pm$11 \\
237.5\tablenotemark{d} & $\lesssim$11 & 162$\pm$37 \\
338.8\tablenotemark{e} & $\lesssim$1 & 200.75$\pm$20 %\\
\enddata
\tablenotetext{a}{The 10, 15, and 33 GHz measurements are considered as the peak intensities in the Briggs Robust$=$0 weighted JVLA images. For ALMA observations, we provide the upper limits.}\vspace{-0.2cm}
\tablenotetext{b}{The integrated flux density measured with aperture photometry.}\vspace{-0.2cm}
\tablenotetext{c}{Spatially unresolved SMA measurements from \citet{Andrews2013}. We assume nominal 10\% (1-$\sigma$) absolute flux uncertainty. }\vspace{-0.2cm}
\tablenotetext{d}{The off-axis measurement given by the archival ALMA Band 6 observations (2015.1.00392.S; \citealt{akeson2019}).}\vspace{-0.2cm}
\tablenotetext{e}{The ALMA Band 7 data published in \citet{Hashimoto2021zztauirs}, quoted with a nominal 10\% (1-$\sigma$) absolute flux error.}\vspace{-0.2cm}
\end{deluxetable}

\subsection{SED modeling}\label{sub:model}

We adopted the strategy of \citet{Liu2019c,Liu2021a} to model the observed flux densities as the contributions of various dust and free-free emission components:
\begin{equation}\label{eqn:multicomponent}
    F_{\nu} = \sum\limits_{i} F_{\nu}^{i} e^{-\sum\limits_{j}\tau^{i,j}_{\nu}}, 
\end{equation}
where $F_{\nu}^{i}$ is the flux density of the dust or free-free emission component $i$ and $\tau^{i,j}_{\nu}$ is the optical depth of the emission component $j$ that obscures the emission component $i$. Because of the limited number of observational constraints (Table \ref{tab:fluxdensity}), we assume that there are no obscurations by other components; that is, $\tau^{i,j}_{\nu}=0$.

To model free-free emission, we employed the formulation of \citet{Mezger1967} and \citet{Keto2003} to approximate its spectral profile ($F^{\rm ff}_\nu$) and optical depth ($\tau^{\rm ff}_\nu$) as follows:
\begin{equation}\label{eqn:flux-ff}
    F^{\rm ff}_\nu = \Omega_{\rm ff} (1-e^{\tau^{\rm ff}_\nu})B_\nu(T_e), 
\end{equation}
\begin{equation}\label{eqn:tau-ff}
    \tau^{\rm ff}_\nu = 8.235 \times 10^{-2}\left(\frac{T_e}{\rm K}\right)^{-1.35}\left(\frac{\nu}{\rm GHz}\right)^{-2.1}\left(\frac{\rm EM}{\rm pc\ cm^{-6}}\right). 
\end{equation}
$B_\nu(T_e)$ is the Planck function. Assuming that all the emissions at the stellar location in our JVLA observations are from free-free emissions, our model requires one component described by an election temperature $T_{e}=$6000 K, emission measure $EM=3\times10^{7}$ cm$^{-6}$\,pc, and solid angle $\Omega_{\mbox{\scriptsize free-free}}=1.25\times10^{-14}$ sr$^{-1}$. The results are presented in Figure~\ref{fig:sedmodel}(a).

The flux density of dust emission ($F^{\mbox{\scriptsize dust}}_{\nu}$) were evaluated based on the analytic radiative transfer solutions published in \citet{Birnstiel2018}.
The free parameters are solid angle ($\Omega_{\rm dust}$), dust temperature ($T_{\rm dust}$), dust mass surface density ($\Sigma_{\rm dust}$), and the maximum grain size ($a_{\rm max}$).
Note that the opacity is determined by the maximum size of the dust grains $a_{\rm max}$ when other grain properties, such as composition and porosity, are fixed. In this study, we quoted the default (i.e., water-ice-coated) DSHARP opacity from \citet{Birnstiel2018}, including absorption and scattering (refer to the following). To model dust thermal emission, we examined these four free parameters (i.e., $a_{\mbox{\scriptsize max}}$, $T_{\mbox{\scriptsize dust}}$, solid angle $\Omega_{\mbox{\scriptsize dust}}$, and column density $\Sigma_{\mbox{\scriptsize dust}}$) interactively instead of fitting them because they are under-constrained by the measurements presented in Table \ref{tab:fluxdensity}.
However, these measurements occur in a range such that meaningful qualitative discussion is plausible.

Regarding opacity, we assume that the dust grain size ($a$) distribution $n(a)$ follows $a^{-q}$ between the minimum and maximum grain sizes ($a_{\mbox{\scriptsize min}}$, $a_{\mbox{\scriptsize max}}$).
It is assumed $n(a)=0$ for $a<a_{\mbox{\scriptsize min}}$ and  $a>a_{\mbox{\scriptsize max}}$.
We also assumed a nominal $a_{\mbox{\scriptsize min}}=10^{-4}$~mm. We examined the probable $a_{\mbox{\scriptsize max}}$ values interactively. The dust spectral profile was not sensitive to $a_{\mbox{\scriptsize min}}$.
We adopted dust temperature $T_{\mbox{\scriptsize dust}}=$23.2 K for all dust components, which is the median of the expected dust temperature in the $r=$45 au ring (\citealt{Hashimoto2021zztauirs}).
Following \cite{Liu2019c}, we approximated small elements in the dust emission sources as geometrically thin slabs, whose spectral profiles can be approximated by Equations 10--20 of \citet{Birnstiel2018}. Note that as mentioned above, since dust opacity also depends on grain properties such as composition \citep[e.g.,][]{Birnstiel2018}, this is a caveat of our dust modeling.

The spectral index $\alpha$ constrained by the ALMA 338.8 GHz observations and the SMA 225.5 GHz observations (Table \ref{tab:fluxdensity}) was lower than 2.0. This low $\alpha$ value favors that the sub-millimeter dust emission in the dust ring of ZZ~Tau~IRS is dominated by optically thick dust (namely, the {\it Ring} component; Table \ref{tab:dustmodel}; Figure \ref{fig:sedmodel}) with $a_{\mbox{\scriptsize max}}\sim$60 $\mu$m. This value is estimated by fitting the sub-millimeter emissions without considering the free-free emissions from the ionized gas. This value is also comparable to the lower limit of an $a_{\mbox{\scriptsize max}}$ estimated in other systems (e.g., 70~$\mu$m $<a_{\rm max}<$ 270~$\mu$m in HL~Tau; \citealp{Kataoka2016b}). Such an $a_{\mbox{\scriptsize max}}$ value suggests that the dust albedo increases with frequency, which helps lower the $\alpha$ value. If this dust albedo effect is insignificant, an unrealistically low dust temperature must be considered (\citealt{Liu2019b}). 

We find that the assumed free-free emission and dust ring cannot account for the total observed flux densities at 8--37 GHz in Figure~\ref{fig:sedmodel}(b), especially at the data point of 15~GHz. This is because that excess of dust thermal emissions occurs when the grain size is comparable to the observed wavelengths \citep[e.g.,][]{Draine2006}. Thus, it is hard to reproduce the SED with centimeter excess by only sub-mm dust grains, and vice versa. If we increase $\Sigma_{\mbox{\scriptsize dust}}$ in the ring to reach the observed 10 and 15~GHz flux densities, the 33 GHz flux density of the dust ring will largely exceed the JVLA measurement. Meanwhile, though opacity of the different grain models in Figure~6 and 11 of \citet{Birnstiel2018} does not show excess at $\lambda \gtrsim$~100~$\mu$m, unknown dust opacity with excess at 15~GHz may explain the SED with excess. Overall, to reproduce the SED, we include another dust component (namely, the {\it crescent} component with different grain size and same opacity; Table \ref{tab:dustmodel}; Figure \ref{fig:sedmodel}), which has a much higher $\Sigma_{\mbox{\scriptsize dust}}$ and much smaller $\Omega_{\mbox{\scriptsize dust}}$ than the dust ring. The two dust components are also naturally motivated by the ALMA dust image of ring with crescent structure \citep{Hashimoto2021zztauirs}. 

To interpret the 10--15 GHz emission with the two dust components, the lower limit of $a_{\mbox{\scriptsize max}}$ in the crescent was $\sim$1.7 mm. A fainter free-free component leads to larger $a_{\mbox{\scriptsize max}}$. Note that the flux of the free-free component in Table~\ref{tab:fluxdensity} is the upper limit. However, it may not be physical to assume a much fainter free-free emission component given that strong radio emission was resolved inside the dust cavity (Figure \ref{fig:JVLAimg}, top row). Therefore, we conclude that grain growth to the mm-sized grains in crescent has occurred, relative to sub-micron-sized interstellar medium. 

\begin{deluxetable}{ lllll }
\tablecaption{
Dust models for ZZ~Tau~IRS\label{tab:dustmodel}
}
\tablewidth{700pt}
\tabletypesize{\scriptsize}
\tablehead{
\colhead{Component} &
\colhead{$T_{\mbox{\scriptsize dust}}$} &
\colhead{$\Sigma_{\mbox{\scriptsize dust}}$} &
\colhead{$\Omega_{\mbox{\scriptsize dust}}$\tablenotemark{a}} &  
\colhead{$a_{\mbox{\scriptsize max}}$} \\
&
\colhead{(K)} &
\colhead{(g\,cm$^{-2}$)} &
\colhead{($10^{-13}$ sr)} & 
\colhead{(mm)} \\
} 
\startdata
Ring & 23 &   9.5  &   33  &   0.06  \\
Crescent & 23 &   180  &   3   &   1.7    \\
\enddata
\tablenotetext{a}{1 sr $\sim$4.25$\times$10$^{10}$ square arcsecond.}\vspace{-0.2cm}
%\tablenotetext{b}{This is the median expected temperature in the ring proposed by \citet{Hashimoto2021zztauirs}.}\vspace{-0.2cm}
%\tablenotetext{c}{This maximum grain size is an upper limit.}
\end{deluxetable}

With a limited number of observational constraints (Table \ref{tab:fluxdensity}), we still cannot determine whether the crescent is obscured by the ring component at high frequencies.
Our fiducial model (Table \ref{tab:dustmodel}) did not consider this obscuration.
If the crescent component is obscured by the ring component at $\gtrsim$200 GHz frequency, the $a_{\mbox{\scriptsize max}}$ value of the ring component needs to be lowered to suppress the attenuation due to dust scattering.
Therefore, the $a_{\mbox{\scriptsize max}}$ value of the ring component in Table \ref{tab:dustmodel} may be regarded as the upper limit (i.e., $a_{\mbox{\scriptsize max}} \lesssim$~60~$\mu$m). 

Existing observations do not rule out the probabilities of larger dust (e.g., $a_{\mbox{\scriptsize max}}\gtrsim$1 cm).
However, a larger dust component might not dominate the flux densities observed at 8--340 GHz possibly due to the much smaller $\Omega_{\rm dust}$ and/or the much lower $\Sigma_{\rm dust}$.

\section{Discussions} \label{sec:discussion}

We compare our results with other systems, including both VLM and more massive stars, in Section~\ref{sub:comp_obs} and discuss the implications of grain growth and plant formation around ZZ~Tau~IRS (\S~\ref{sub:implication}).

\subsection{Comparison with other centimeter observations}\label{sub:comp_obs}

T~Tauri stars and intermediate-mass stars were observed at centimeter wavelengths. Multi-frequency radio observations have been performed to distinguish dust thermal emissions from free-free emissions of ionized jets, and multi-frequency radio observations have been performed to construct SEDs \citep[e.g.,][]{Testi2003a,Perez2012,Perez2015a,Tazzari2016a}. Although the spectral index of free-free jets is less than one (e.g., \citealp{Anglada1996}, refer also to \S~\ref{sub:model}), dust thermal emissions, including dust self-scattering, exhibit a spectral index of more than one \citep{Miyake1993,Zhu2019b,Liu2019b}. The change in the slope of SEDs is attributed to different emission mechanisms. Using disk emission models, the dust grain size has been estimated in these disks \citep[e.g.,][]{Testi2003a,Perez2012,Perez2015a,Tazzari2016a}. However, recent studies on dust scattering in disks suggest that dust scattering reduces thermal emissions from the disk \citep{Zhu2019b,Liu2019b}. The modeling to estimate $a_{\mbox{\scriptsize max}}$ needs to incorporate the effect of dust scattering. Some modelings incorporating dust scattering suggest that $a_{\mbox{\scriptsize max}}$ is larger than the centimeter size in the ring around HD~163296 \citep{Guidi2022a}, although the modeling contains large uncertainties.

Regarding brown dwarfs and VLM stars, though centimeter wavelength observations have been conducted  \citep[e.g.,][]{Osten2006a,Rodriguez2017a,zapa17,Alves2020a,Greaves2022a}, most of them have not been detected. One source of a young brown dwarf FU~Tau~A was detected in the 3.0 cm emission \citep{Rodriguez2017a}. However, because the reported beam size is larger than $\gtrsim$1$\arcsec$, the emission is not spatially resolved. The typical disk size around VLM stars is less than 1$\arcsec$ \citep[e.g., ][]{Kurtovic2021a}, implying a similar or smaller size of the brown dwarf disks. Thus, it is difficult to distinguish dust thermal emissions from the free-free emissions of ionized jets. Similar to the case of more massive stars, constructing SEDs using multiple radio wavelength observations helps to distinguish different components \citep[e.g.,][]{Testi2003a,Rodmann2006a}. Future multiple-wavelength radio observations with high sensitivity will shed light on whether dust grains grow to larger sizes around the brown dwarfs and VLM stars.

\subsection{Implication of planet formation around ZZ~Tau~IRS}\label{sub:implication}

Crescent structures are often interpreted as dust trapping vortices, which serve as an ideal forming site of planetesimals by efficiently accumulating millimeter/centimeter grains \citep[e.g.,][]{Barge1995}. When local dust densities are high enough, planetesimals may form by gravitational instability \citep{joha2014a}. Subsequent pebble accretion eventually led to the formation of planets \citep{Liu2020b}. The pebble accretion model predicts that larger pebbles (centimeter size) accelerate planet formation by being efficiently accreted onto planet cores \citep{Morbidelli2015,liu2018,Ormel2018}. Consequently, in crescent with $a_{\mbox{\scriptsize max}}$ of millimeter size (Table~\ref{tab:dustmodel}), planet embryos may grow more efficiently than other regions with sub-millimeter-sized grains. 
While the pebble accretion scenario prohibits giant planet formation around VLM stars \citep{liu2019a,Liu2020a}, larger values of disk-scale height might circumvent this difficulty. This is because the inner wall of the ZZ~Tau~IRS ring is directly irradiated by the central star, resulting in the heating of the ring. Thus, the disk scale height at the ring might increase. Future near-infrared polarimetric observations and CO gas observations tracing the disk surface are necessary to constrain the disk-scale height.

\section{Conclusion} \label{sec:conclusion}

We observed the VLM star ZZ~Tau~IRS with JVLA at Ka (33~GHz), Ku (15~GHz), and X (10~GHz) bands to search for centimeter-sized pebbles in its ring with crescent. We detected significant signals from the ring with crescent at 33~GHz and 15~GHz, whereas only the point source attributed to free-free emission was detected at 10~GHz. These are the first robust detections of centimeter emissions from the disk around the VLM star.

The 8--340 GHz emission was interpreted by a simple SED model with three components though our modeling efforts rely on uncertain dust properties: a nominal free-free emission component, dust ring component with $a_{\mbox{\scriptsize max}}\lesssim$60~$\mu$m, and dust crescent component with $a_{\mbox{\scriptsize max}}\gtrsim$1~mm. The grain growth in the ZZ~Tau~IRS disk proceeds to (sub-)millimeter-sized grains, relative to sub-micron-sized interstellar medium.

Pebble accretion models predict that larger grains accelerate planet formation. Our case studies of ZZ~Tau~IRS suggest that planet formation in the crescent containing millimeter-sized grains might proceed more efficiently than in other regions with sub-millimeter-sized grains. Observations of VLM stars at centimeter wavelengths are limited, and the detection of centimeter wavelengths from disks around VLM stars is only ZZ~Tau~IRS. Hence, it is unclear whether pebbles in disks around VLM stars are commonly limited to (sub-)millimeter-sized grains or grow to centimeter or larger pebbles. Increasing the sample size of sensitive centimeter observations is necessary to understand grain growth and planet formation in disks around VLM stars. 

\begin{acknowledgments}

The authors thank the anonymous referee for a timely and constructive report. The National Radio Astronomy Observatory is a facility of the National Science Foundation, operated under a cooperative agreement by Associated Universities, Inc.
This study used the following ALMA data: ADS/JAO. ALMA \#2015.1.00392.S and 2016.1.01511.S. 
ALMA is a partnership of ESO (representing its member states), NSF (USA), and NINS (Japan), along with NRC (Canada), MOST and ASIAA (Taiwan), and KASI (Republic of Korea), in cooperation with the Republic of Chile. The Joint ALMA Observatory is operated by ESO, AUI/NRAO, and NAOJ.
H.B.L. was supported by the National Science and Technology Council (NSTC) of Taiwan (Grant Nos. 108-2112-M-001-002-MY3, 111-2112-M-001-089-MY3, and 110-2112-M-001-069). This study was supported by JSPS KAKENHI Grant Numbers 21H00059 and 22H01274. R.D. acknowledges the support from the Alfred P. Sloan Foundation and the Natural Sciences and Engineering Research Council of Canada.
%We would like to thank Editage (www.editage.com) for English language editing.

\end{acknowledgments}

\software{
          astropy \citep{2013A&A...558A..33A},  
          Numpy \citep{VanDerWalt2011}, 
          CASA \citep[v6.4.3; ][]{mcmu07},
          }

\facilities{JVLA, ALMA}

\bibliography{sample63}{}
\bibliographystyle{aasjournal}

\end{document}